\documentclass[amsmath,amssymb,reprint,nobibnotes]{revtex4-2}
\usepackage{graphicx}
\usepackage{bm}
\usepackage{amsmath} 
\usepackage{graphics}
\usepackage[dvipsnames]{xcolor}
\usepackage[export]{adjustbox}
\usepackage[mathlines]{lineno}
\usepackage{braket}
\usepackage[compact]{titlesec}
\begin{document}
\title{Continuous and reversible electrical-tuning of fluorescent decay rate \\ via Fano resonance}
\author{Emre Ozan Polat$^{\bf (1,2,6,+)}$}\email{emre.polat@bilkent.edu.tr}
\author{Zafer Artvin$^{\bf (3,6)}$}
\author{Yusuf Şaki$^{\bf (3,6)}$}
\author{Alpan Bek$^{\bf (3,6)}$}
\author{Ramazan Sahin$^{\bf (4,5,6,+)}$}\email{rsahin@itu.edu.tr}%

\affiliation{${\bf (1)}$ {Department of Physics, Bilkent University, 06800 Ankara, Türkiye}}
\affiliation{${\bf (2)}$ {UNAM- National Nanotechnology Research Center and Institute of Materials Science and Nanotechnology, Bilkent University, 06800 Ankara, Türkiye}}
\affiliation{${\bf (3)}$ {Department of Physics, Middle East Technical University, 06800 Ankara, Türkiye}}
\affiliation{${\bf (4)}$ {Department of Physics, Akdeniz University, 07058 Antalya, Türkiye}}
\affiliation{${\bf (5)}$ {Türkiye National Observatories, TUG, 07058, Antalya, Türkiye}}
\affiliation{${\bf (6)}$ {Institute of Nuclear Sciences, Hacettepe University, 06800 Ankara, Türkiye}}
\affiliation{${\bf (+)}$ {These two authors contributed equally to this work.}}

\date{\today}

\begin{abstract}
We demonstrate that the decay rates of a fluorescent molecule can be controlled by electrically shifting a transparency introduced by a Fano resonance. An auxiliary quantum object (QO), located at the hotspot of a plasmonic nanoparticle, suppresses plasmonic excitation at its level spacing $\omega_{\rm QO}$. As a result, the local density of states (LDOS) associated with the plasmonic spectrum is also suppressed at $\omega = \omega_{\rm QO}$. By shifting $\omega_{\rm QO}$ via an applied voltage, we continuously tune the radiative and nonradiative decay rates of the fluorescent molecule by up to two orders of magnitude. This mechanism offers a valuable tool for integrated quantum technologies, enabling on-demand entanglement and single-photon sources, voltage-controlled quantum gate operations, and electrical control of superradiant-like phase transitions. The approach also holds promise for applications in super-resolution microscopy and surface-enhanced Raman spectroscopy (SERS).
\end{abstract}

\maketitle

\section{Introduction}
Electromagnetic interaction between an atomic system and a closely located metal nanostructure (MNS) has provided a viable platform for researchers to investigate the nature of nanophotonic phenomena yielding series of notable innovations such as SERS, that allows for single molecule detection \cite{serafinelli2022plasmonic}, single photon transistors realizing strong nonlinear interactions at the single-photon level \cite{chang2007single}, and cavity-free subwavelength confinement of optical fields using 0D-1D structures \cite{akimov2007generation}.  At the heart of the demonstrated innovations lies the Purcell effect \cite{purcell1995spontaneous} which defines the change in the lifetime of an atomic/molecular excited state. It takes place due to the modifications in the number of available optical states (local density of optical states, LDOS) emitter can couple. 

This, in turn, results in the enhancement or suppression of the spontaneous emission rate of the emitter. The effect is prominent especially when the molecular system is located in a resonant cavity  \cite{vahala2003optical} or near a MNS  \cite{lakowicz2005radiative} that modifies the electromagnetic environment. 
The modified decay rate is given by the change in the LDOS with respect to free space, such as $\gamma$=$(LDOS)_{\rm env}$/$(LDOS)_{\rm free}$ $\times$ $\gamma_0$ where $\gamma_{0}$ is the free space decay rate. In the presence of an absorptive material, such as an MNS, enhanced molecular decay rate can take place both as photon emission and nonradiative photon transfer (absorption) into the MNS ---apart from the intrinsic nonradiative losses within the emitter. The former demonstrates itself as fluorescence enhancement \cite{camposeo2015metal} while the latter is linked to optical losses. 

Active control of Purcell effect is a key to enable switchable light emitting nanophotonic devices which represents a fast-growing field due to its direct contribution to telecommunication and integrated quantum technologies \cite{lu2017dynamically,munnix2009modeling,bimberg2009quantum}. While there exist studies on the control of spontaneous emission via Purcell effect, they mostly rely on altering the cavity size/parameters representing a passive control scheme \cite{canet2012purcell,bennett2005microcavity,moritake2016emission}. Cavity-dependent demonstrations of Purcell effect have attracted great attention due to the ability to tune the fluorescence emission spectra of molecules. However, it comes with certain limitations for active control of Purcell effect such as fixed cavity components and mirror spacings \cite{steiner2008controlling}. Although this limitation was partially solved within a tunable microcavity scheme, the applicability of the approach is still limited in the context of integrated circuits due to the requirement of immersion oils and CMOS incompatible elements \cite{chizhik2009tuning}.

Alternatively, some active methods to switch photoluminescence (PL) have also been demonstrated using various physical phenomena such as quantum confined stark effect\cite{dots1997quantum,xu2008manipulating}, electric field induced quenching  \cite{bozyigit2013origins}, Förster resonant energy transfer (FRET) \cite{salihoglu2016graphene}, and implementation of electroactive polymers in redox reaction  \cite{chen2012electric,wang2013tuning}. Although it was shown to be possible to switch fluorescence actively, these methods utilize quenching in order to turn off the fluorescence. That is, molecular excitation is almost completely transferred (lost) into the MNS and not kept in the emitter. Moreover, the modulation depths are poor and some of the studies report only a limited number of successive modulations due to hysteresis in the cyclic performance prohibiting a complete recovery between on and off states \cite{salihoglu2016graphene,chen2012electric,wang2013tuning}. Furthermore, the response time of those applications reaches maximum millisecond regime which is not compatible with CPU clock speed range (GHz) and barely supports current display technology with frame rates (FPS) of 60-144 Hz. Despite the attempts to use reverse bias providing 200:1 contrast ratio and 300 ns response time \cite{xie2022voltage}, the usage of extremely high (3$\times$10$^6$ V/cm$^2$) external field prohibits the technological use of reported methodologies.

Therefore, achieving continuous and reversible control within a compact device geometry that provides large modulation depths with (and over) CPU clock speed compatibility remains as a great challenge in active control of fluorescence for technology applications. While certain progresses have been achieved on the electrical control of the Purcell effect~\cite{lu2017dynamically,munnix2009modeling,bimberg2009quantum,canet2012purcell,bennett2005microcavity,moritake2016emission,steiner2008controlling,chizhik2009tuning,dots1997quantum,xu2008manipulating,bozyigit2013origins,salihoglu2016graphene,chen2012electric,wang2013tuning,xie2022voltage}, they either remain below the CPU clock speeds or possess poor modulation depths.

In this paper, we introduce a new method for fast electrical tuning of the Purcell effect that also achieves large modulation depths. Unlike conventional methods used in the literature~\cite{lu2017dynamically,munnix2009modeling,bimberg2009quantum,canet2012purcell,bennett2005microcavity,moritake2016emission,steiner2008controlling,chizhik2009tuning,dots1997quantum,xu2008manipulating,bozyigit2013origins,salihoglu2016graphene,chen2012electric,wang2013tuning,xie2022voltage}, our approach relies on tuning a transparency in the plasmonic response that is introduced via a Fano resonance~\cite{leng2018strong,limonov2017fano,tacsgin2018fano}. Our method outperforms existing electric-control techniques by offering a modulation depth of approximately 200 (i.e., 20,000\%) and a response time on the order of a picosecond. It allows continuous tuning of the radiative and nonradiative decay rates of a fluorescent molecule, or the control of spontaneous parametric down-conversion in quantum integrated circuits.

An auxiliary quantum object (QO) —such as a quantum dot or a collection of defect centers~\cite{murata2011high,rotem2007enhanced,ngandeu2024hot}— is placed at the hotspot of a core-shell nanoparticle~(CSNP) and introduces the Fano resonance (i.e., a transparency)~\cite{leng2018strong,limonov2017fano,tacsgin2018fano}, as shown in Fig.~1. This transparency appears at the level-spacing of the QO, which is $\omega_{\rm \scriptscriptstyle QO}$ = 766 nm in our simulations~\footnote{We refer the frequencies by the corresponding free-space wavelength, governed by $\omega=c/\lambda$. }. The transparency suppresses plasmonic excitation and, in turn, eliminates the plasmonic enhancement in the local density of states~(LDOS) at $\omega = \omega_{\rm \scriptscriptstyle QO}$~(see Fig.~2). When the metastable transition frequency of a fluorescent molecule $\omega_{\rm \scriptscriptstyle FM}$ —located 15 nm away— matches the QO level spacing (i.e., $\omega_{\rm \scriptscriptstyle FM} = \omega_{\rm \scriptscriptstyle QO}$), the plasmonic enhancement of fluorescence is reset to its vacuum value, $\gamma_r = \gamma_0$ (compare solid and dashed lines in Fig.~2).

The QO level spacing—and hence the spectral position of the transparency—can be controlled by an applied voltage~(see Fig.~3a). For example, if $\omega_{\rm \scriptscriptstyle QO}$ shifts to $\omega_{\rm \scriptscriptstyle QO}(V)$ = 767 nm, the new transparency appears at 767 nm. As a result, the LDOS associated with the decay at $\omega_{\rm \scriptscriptstyle FM} = 766$ nm changes from $\gamma_r = \gamma_0$ to $\gamma_r(V)=10\gamma_0$. By continuously shifting the QO level spacing, the decay rate can be tuned smoothly (see Fig.~3b and 3c). This method allows full control over the range $\gamma_r = 1$ to $215 \gamma_0$ by adjusting the voltage from $V_{\rm app} = 0$ to 1 V. This mechanism enables our approach to achieve both large modulation depths and fast response times, limited only by the circuit readout rate (GB/s).

We emphasize that the demonstrated system is fundamentally different from previously reported active PL switching and modulation effects that are based on screening of the PL for an external observer.
 By trapping the fluorescent state in the molecule (i.e., not quenching it), the nature of our switching process exhibits a crucial property with the ability to hold and release the interested fluorescent state by voltage application.

\begin{figure}
\includegraphics[width=0.48\textwidth]{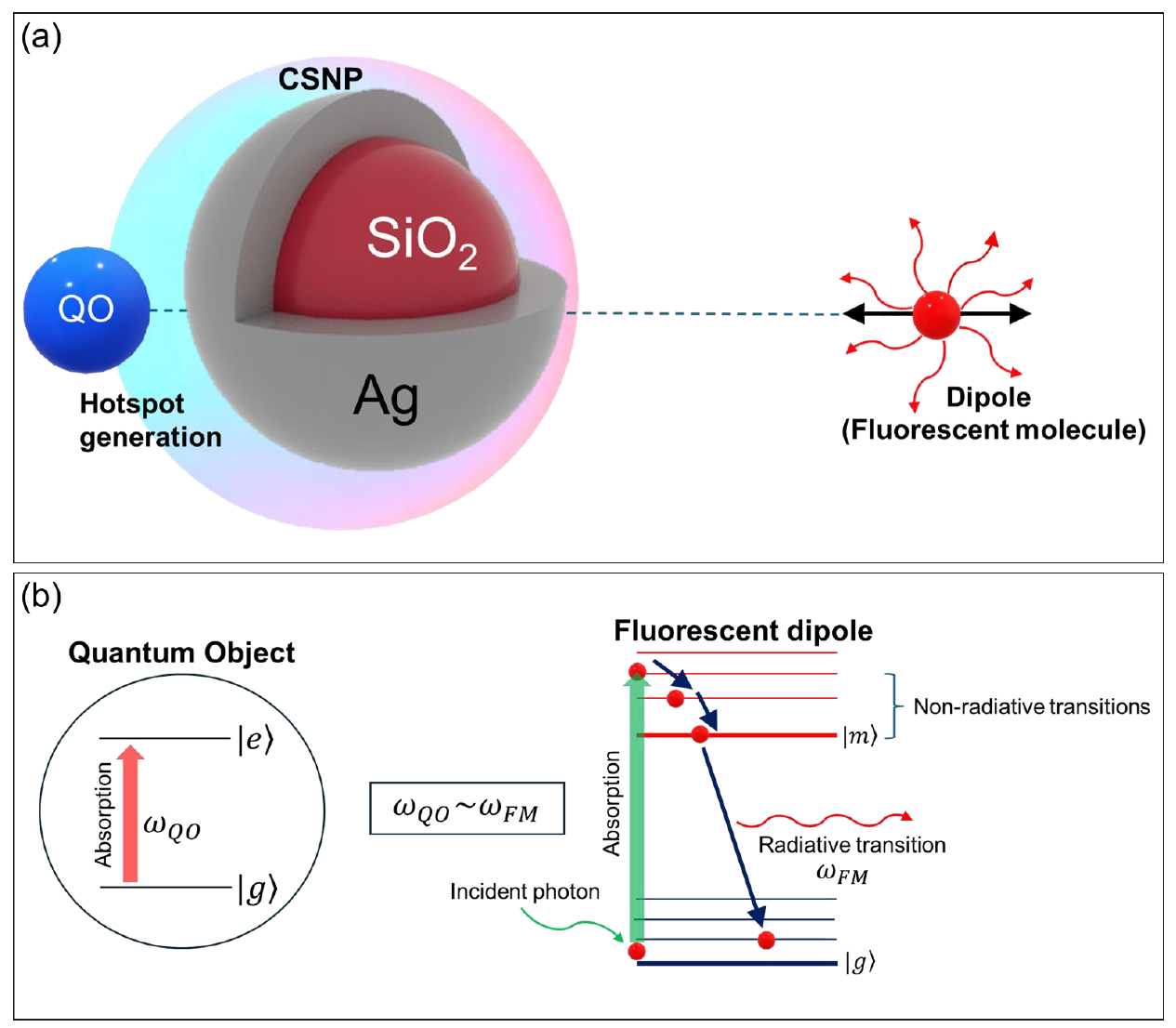}
\caption{\label{fig1} Proof-of-principle illustration of the phenomenon on a sample configuration. (a) An auxiliary QO, representing a dense collection of defect-centers or densely ion-planted G-centers is positioned at the hotspot. FM (dipole) whose decay rates are of interest is located at 15 nm distance. (b) The two-level QO and the Jablonski diagram of the FM. The absorption frequency of the QO ($\omega_{\rm \scriptscriptstyle QO}$) and radiative transition of the dipole ($\omega_{\rm \scriptscriptstyle FM}$) are similar. The QO introduces a Fano transparency in the spectrum of the CSNP that controls the CSNP-induced LDOS.}
\end{figure}

We provide a proof-of-principle demonstration of active and reversible tuning of both radiative and nonradiative decay rates of a fluorescent molecule~(FM) via exact solutions of the 3D Maxwell equations. We perform our simulations using in Lumerical --a reliable FDTD simulation toolbox~\cite{Ansys2024}.
In our demonstrated system, the presence of a plasmonic SiO$_2$/Ag core (SiO$_2$) shell (Ag) nanoparticle (CSNP) boosts the LDOS of a FM (Fig. 1a). When an auxiliary QO, such as a quantum dot (QD) or a defect-center of resonance $\omega_{\rm \scriptscriptstyle QO}$,  is positioned near the CSNP, it introduces a Fano transparency at $\omega_{\rm \scriptscriptstyle QO}=\omega_{\rm \scriptscriptstyle FM}$  \cite{wu2010quantum}. Fano transparency is the plasmon-analog of electromagnetically-induced-transparency (EIT)  \cite{fleischhauer2005electromagnetically}. Coupling of a dark mode or a QO to a bright plasmon mode introduces two paths for the absorption \cite{shah2013ultrafast}. When the two paths work out of phase, absorption is canceled, and a transparency window appears in the spectrum of the MNS \cite{alzar2001classical}. Therefore, an absorption and scattering dip appears at $\omega_{\rm \scriptscriptstyle QO}$. In Sec.~I of the Supplementary Material~(SM)~\cite{SM}, we explain why and how such a transparency arises in the plasmonic excitation spectrum, using a single equation with detailed discussion. 

In this work, by choosing $\omega_{\rm \scriptscriptstyle QO}$  about the transition frequency $\omega_{\rm \scriptscriptstyle FM}$  of the FM’s metastable (fluorescence) state (Fig. 1b), we show that the presence of the QO turns off the LDOS that used to be enhanced due to the presence of the CSNP (Fig. 2a). In other words, at $\omega_{\rm \scriptscriptstyle QO}=\omega_{\rm \scriptscriptstyle FM}$ turns off the extra LDOS occurring due to the presence of the CSNP. Additionally, it also turns off the nonradiative decay rate emerging due to the dissipation at the Ag shell (Fig. 2b). 

More interestingly, as the QO’s resonance can be shifted \cite{larocque2024tunable,miller1985electric} via applied voltage (Fig. 3a), the decay rate patterns of Figs. 2a and 2b are also shifted electrically, enabling the continuous tuning of the decay rates (Figs. 3b and 3c).  Thus, the LDOS, which is initially enhanced due to the CSNP, can also be tuned continuously via the applied voltage within sub-picosecond response time. We emphasize that a ~20 meV resonance shift can tune the radiative decay rate between $\gamma_{\rm r}$=1--215$\gamma_0$  continuously. 

This method, being reversible and cavity-free, allows for nanofabrication approaches for the integration in quantum circuits and conventional Si-based readout circuits (see the Supplementary Material, SM). Compared to active electroluminescent counterparts, Fano resonance provides an incomparably large range of tuning corresponding to 215 times (21500\%) modulation. It presents technologically compatible operation frequencies since the response time of the Fano resonance is of the order of picosecond and therefore the tuning rate is only limited by the speed of the solid-state device. We believe that the demonstrated method manifests itself as an invaluable tool that enables so-far unrealized implementations ranging from voltage-controlled superpositions of hyperfine states to on-demand single-photon imaging techniques.
\section{Results}

Fig. 1a shows a sample configuration we have simulated as a proof-of-principle demonstration of the phenomenon. In our simulations we use a SiO$_2$/Ag CSNP of diameter 105 nm where plasmonic losses limiting the use of MNSs in optoelectronics are compensated via optical gain incorporated in the core insulator (SiO$_2$) region (100 nm in diameter)  \cite{rajput2020forster,issah2023epsilon}. The Ag-shell is a 5 nm thick layer that is deposited uniformly on the SiO$_2$ core. An auxiliary QO of 40 nm diameter representing a densely implanted G-centers (or nitrogen-vacancy (NV) defect centers) \cite{murata2011high,rotem2007enhanced,ngandeu2024hot} is located 5 nm to the Ag-shell and introduces the Fano transparency at $\omega_{\rm \scriptscriptstyle QO}$. FM (dipole), polarized along the x-axis, is positioned on the opposite side at 15 nm distance. We calculated the LDOS for the radiative and nonradiative decay rates of the dipole (fluorescent molecule FM) using finite difference time domain method (FDTD)  \cite{Ansys2024}. In the demonstrated configuration, we first show that Fano transparency can switch off the initially enhanced emission (radiative decay) rate of the FM. Fig. 2a shows the radiative decay spectrum of the dipole (FM) with and without the QO in the system. The QO residing at the hotspot of the CSNP turns off the CSPN induced enhancement in the LDOS. Decay rate drops from 215$\gamma_{0}$ to $\gamma_0$ at 
 $\omega_{\rm \scriptscriptstyle QO}=\omega_{\rm \scriptscriptstyle FM}$. For other fluorescence wavelengths, decay rate suppression is weaker. One should note that decay of a molecule from its excited state always includes a quantum probabilistic nature \cite{yucel2020single}. The $\gamma_{0}$ is to be understood as an expectation value.


 \begin{figure}[ht]
\includegraphics[trim={0.6cm 0 0.0cm 0cm},clip, width=0.48\textwidth] {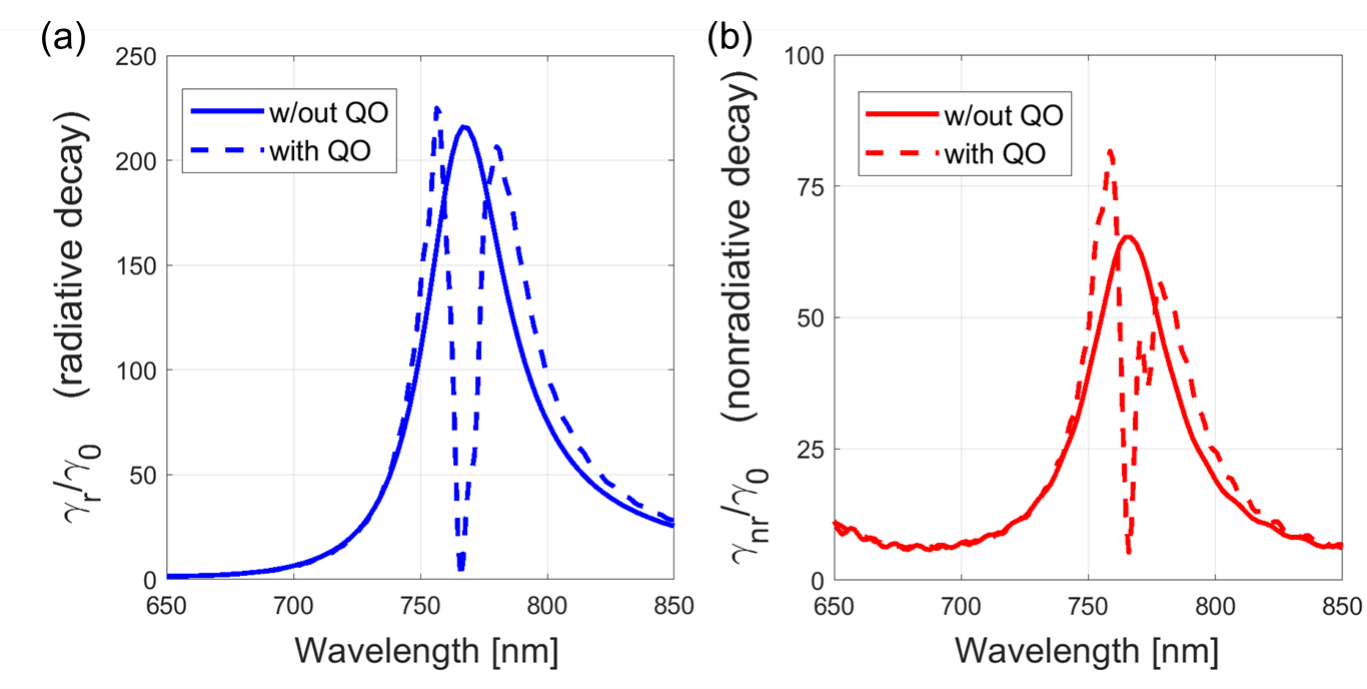}
\caption{\label{fig2} Suppression of the (a) radiative ($\gamma_{\rm r}$) and (b) nonradiative ($\gamma_{\rm nr}$) decay rates by Fano resonance. A QO residing at the hotspot can turn off the CSNP-induced LDOS.}
\end{figure}

In Fig. 2b, we observe a similar effect for the nonradiative decay rate. The presence of the QO also suppresses the absorption losses that the CSNP induces initially. The nonradiative decay is suppressed down to 5$\gamma_{0}$ from 65$\gamma_{0}$. We model the QO by a Lorentzian dielectric function  \cite{wu2010quantum,shah2013ultrafast,liu2017electromagnetically} of resonance $\omega_{\rm \scriptscriptstyle QO}$=766 nm, linewidth 
$\sim 10^{10}$ Hz and oscillator strength $f=0.2$. We used the corresponding experimental dielectric functions for CSNP composed of Ag shell and ${\rm SiO_2}$ core materials  \cite{johnson1972optical,palik1998handbook}. To show the promises of our method, we further study the voltage-controlled continuous tuning of the decay rates. Application of an external electric field on the QO shifts the resonance $\omega_{\rm \scriptscriptstyle QO}$ (Fig. 3a). 

Thus, the dashed curves in Figs. 2a and 2b, in particular the dips, also shift. The FDTD simulations show that the radiative and nonradiative decay rates can be tuned between $\gamma_{\rm r}$=1--215$\gamma_{0}$ and $\gamma_{\rm nr}$=5--70$\gamma_{0}$, respectively, in our sample setup (Figs. 3b and 3c). Moreover, as the electrical shift in the resonance is reversible, modulation of $\gamma_{\rm r}$ and $\gamma_{\rm nr}$ are also reversible. In Figs. 3b and 3c, the fluorescence is fixed at $\omega_{\rm \scriptscriptstyle FM}$  = 766 nm and $\omega_{\rm \scriptscriptstyle QO}$ is altered. Only a $\sim 20$ meV tuning of the QO resonance is sufficient for such a control noting that much larger shifts have already been reported even for 1V of applied potential \cite{shibata2013large}. 

\begin{figure}
\includegraphics[width=0.48\textwidth]{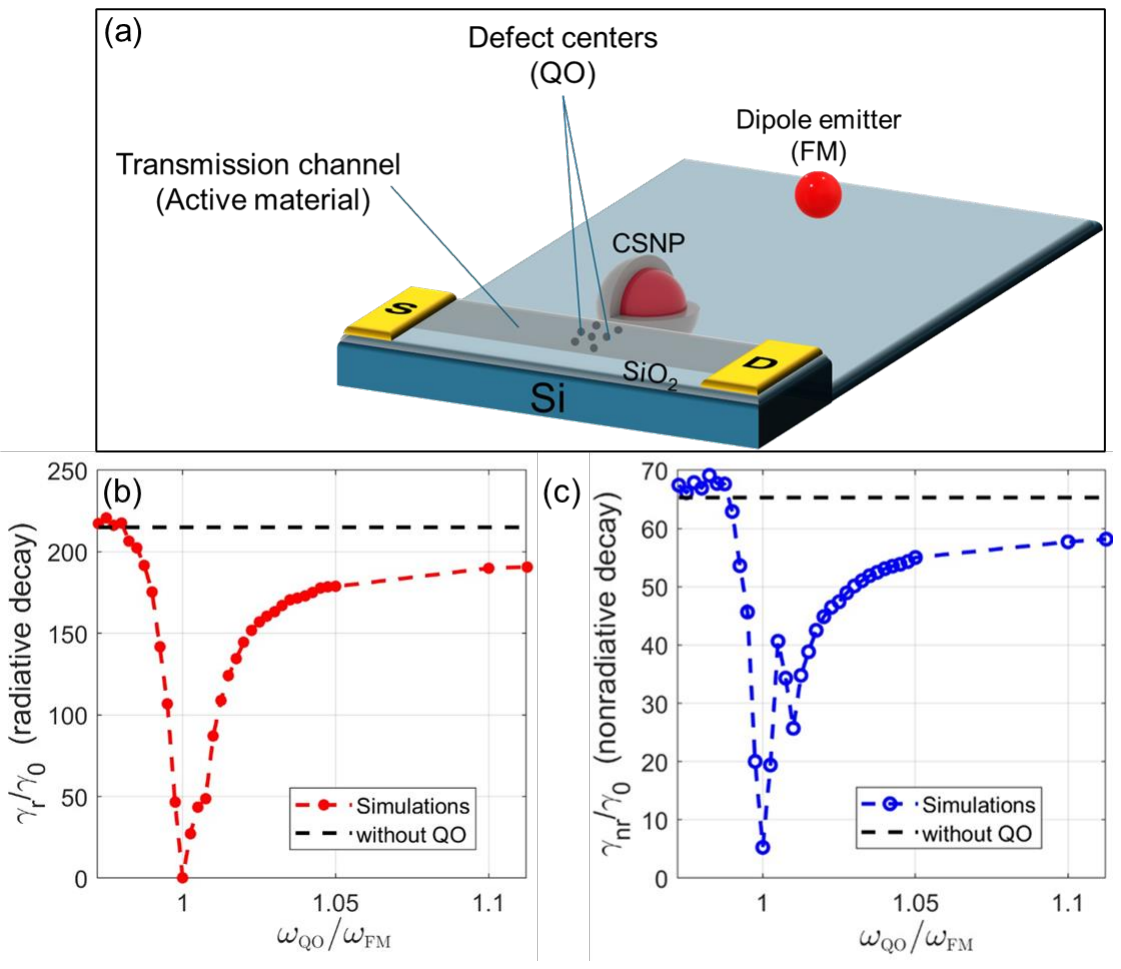}
\caption{\label{fig3} Continuous electrical tuning of radiative and nonradiative decay rates. (a) Illustration of a possible device scheme for continuous electrical tuning of the decay rate. In a FET structure, the collection of defect centers \cite{murata2011high,rotem2007enhanced} that act as the modeled QO, are biased through source and drain electrodes and gated by Si substrate through the ${\rm SiO_2}$ dielectric. The step-like structure is used to align the center of CSNP to the defect centers. (b) Radiative decay rate of the fluorescent dipole molecule under the influence of electrically controlled Fano transparency. Shift of $\omega_{\rm \scriptscriptstyle QO}$ with respect to $\omega_{\rm \scriptscriptstyle FM}$ changes the efficiency of the suppression for each voltage step, allowing continuous electrical tuning of the decay rate. Only a $\sim$ 20 meV resonance tuning is enough to achieve the continuous control between $\gamma$=1--215$\gamma_{0}$. (c) A similar control over nonradiative decay rate is also possible.}
\end{figure}

Fig. 3a shows the schematic illustration of a possible device where the collection of the defect centers is modelled as the QO. Several experimental methods have been reported for the formation of G-centers in Si/${\rm SiO_2}$ as well as NV centers in diamond, such as the incorporation of carbon rich Si crystals in reactive ion etching process  \cite{rotem2007enhanced}, functionalization of Si surface with organic molecules \cite{murata2011high} and hot ion implementation in diamond \cite{ngandeu2024hot}. In our model, by representing a collection of G-centers as a QO, the external field is provided through the gate of a field-effect transistor (FET), where the active material that contains defect centers are biased through the source-drain channel. To provide the agreement between our model and the FET scheme, a step-like structure is employed merely to provide the spatial alignment of the defect centers to the CSNP. We note that an inhomogeneous broadening among the resonances of, e.g., G-centers results a similar broadening in the Fano transparency dip in Figs. 2a and 2b. In this case, even better control over relatively broader fluorescence spectra can be achieved.

Thus, we numerically demonstrate that we can actively control both the radiative and nonradiative decay rates of a FM for a narrow wavelength regime in a cavity-free environment. Our simulations were originally performed in vacuum therefore the dielectric environment of any possible device scheme (as shown in Figure 3a) would shift the plasmon resonances. This, however, won’t change the nature of the demonstrated phenomenon. Moreover, in the supplementary material (SM) we demonstrate that such a control mechanism works for different choices of the FM-MNP distances (see Sec.~IV of the Supplementary Material~(SM)~\cite{SM}).

\section{Conclusion, Discussion and Outlook}

We present a proof-of-principle numerical demonstration of voltage-tunable decay rates. Unlike conventional methods~\cite{lu2017dynamically,munnix2009modeling,bimberg2009quantum,canet2012purcell,bennett2005microcavity,moritake2016emission,steiner2008controlling,chizhik2009tuning,dots1997quantum,xu2008manipulating,bozyigit2013origins,salihoglu2016graphene,chen2012electric,wang2013tuning,xie2022voltage}, we utilize a transparency introduced by a Fano resonance. This transparency occurs at the level spacing $\omega_{\rm QO}$ of an auxiliary quantum object (QO) placed at the hotspot of a plasmonic nanoparticle (see Fig. 1). We control the decay rate of a fluorescent molecule located 15 nm away from the nanoparticle. The transparency suppresses the LDOS induced by the nanoparticle (CSNP), which in turn reduces the decay rate of the fluorescent molecule. By shifting $\omega_{\rm QO}$ via an applied voltage, we control the position of the transparency and, consequently, the decay rate continuously (see Figs. 3b and 3c). Our simulations—based on exact solutions of the 3D Maxwell equations—demonstrate up to two orders of magnitude modulation in the decay rate.

%
%

The phenomenon addresses a wide range of applications as we are demonstrating a fundamental issue in light-matter interaction, that is electrical-tuning of LDOS. This provides an invaluable tool especially for integrated quantum technologies.The technique offers ultimate control over on-demand single-photon sources and entanglement generation in circuits. The method can also be utilized to control spontaneous parametric down conversion rate by choosing $\omega_{\rm \scriptscriptstyle QO}$ at $\omega/2$.
 These also imply voltage-controlled quantum gate operations. The phenomenon of superradiant phase transition, which depends on the decay rate, now can be electrically turned on/off in an integrated circuit. Thus, normal and superradiant phase transitions between zero and a very strong entanglement state \cite{tasgin2017many},  can be employed to serve as a controllable collective entanglement resource in the integrated circuits. In addition, quantum batteries can become voltage-controlled based on the demonstrated phenomenon ~\cite{alicki2013entanglement,ferraro2018high}.

Change in the LDOS not only alters the decay rates but can also tune coupling of a fixed intensity light to dipoles. This enables electrical tuning of light-matter interaction and stimulated transitions in ion-trap configurations. It is worth noting that the coefficients among two or more hyperfine states can be tuned very quickly without necessitating a change in the pump intensity of different polarizations~\cite{bluvstein2024logical}. For appropriate choice of MNS-shapes (i.e. helical MNSs with spin angular momentum), the phenomenon can be used to enhance/control LDOS selectively among different polarizations. Thus, the method provides a compact, voltage-tunable and fast preparation of single atom/ion states or collective states of their ensembles. For direct applications with the state-of-the-art ion traps, however, one also needs to take extra care of interaction of MNS with the trap potentials, though such potentials result only weak couplings with MNSs. The same features are also valid for the quantum gate operations. 

In this work, we study proof-of-principle demonstration of the phenomenon in free space where the tuning starts at $\gamma_{0}$ and ranges to 215$\gamma_{0}$. It is well studied that decay rates of FMs can be lowered down to values $\gamma_{0}'$ that are much smaller than $\gamma_{0}$, e.g., when they are in photonic crystal cavities~\cite{lodahl2004controlling}. Such setups, in principle, can provide electrical tuning of decay rates in terms of the multiples of much smaller base values $\gamma_{0}'$. Demonstrated device structure can be achieved by various fabrication techniques and device construction schemes. (See the SM for details of reported fabrication steps.) Finally, the effect that we observe can be made more efficient by implementing more optimal setups.

We anticipate that the demonstrated technique can also be utilized in quantum optics for the on-demand generation of single-photons (or photon-pairs) and their synchronization with external systems such as a vibrating scanning probe microscopy (SPM) tip, or a confocal scanning laser microscope. Unlike electrically-triggered quantum dots or defects~\cite{yuan2002electrically}, our method tunes the mean decay time after electrical excitation, and it can also be combined with such sources. Thanks to the ultimate control on the time gated activation of the emitter, the technique can also prove to be useful in performing spin-echo based quantum measurements/sensing applications such as quantum magnetometers~\cite{sturner2021integrated}  or quantum information processing~\cite{madsen2022quantum}.  Alternately, it can also serve as a voltage tunable band rejection filter for the radiative and nonradiative wavelengths falling into the corresponding Fano resonance linewidth. In general, the suitable execution of the demonstrated technique can form a base technology in the applications that require fine tuning of a specific fluorescence spectrum such as SERS and super-resolution microscopy.

\begin{acknowledgments}
We gratefully thank Mehmet Emre Tasgin and Serkan Ates for discussions leading to this research. This research is conducted in the meetings that took place at Institute of Nuclear Sciences, Hacettepe University. E.O.P acknowledges the personal research fund (KAF) of Bilkent University. 
R.S. acknowledges support from TUBITAK No. 121F030 and 123F156.

\end{acknowledgments}

\bibliography{bibliography}

\begin{thebibliography}{52}%
\makeatletter
\providecommand \@ifxundefined [1]{%
 \@ifx{#1\undefined}
}%
\providecommand \@ifnum [1]{%
 \ifnum #1\expandafter \@firstoftwo
 \else \expandafter \@secondoftwo
 \fi
}%
\providecommand \@ifx [1]{%
 \ifx #1\expandafter \@firstoftwo
 \else \expandafter \@secondoftwo
 \fi
}%
\providecommand \natexlab [1]{#1}%
\providecommand \enquote  [1]{``#1''}%
\providecommand \bibnamefont  [1]{#1}%
\providecommand \bibfnamefont [1]{#1}%
\providecommand \citenamefont [1]{#1}%
\providecommand \href@noop [0]{\@secondoftwo}%
\providecommand \href [0]{\begingroup \@sanitize@url \@href}%
\providecommand \@href[1]{\@@startlink{#1}\@@href}%
\providecommand \@@href[1]{\endgroup#1\@@endlink}%
\providecommand \@sanitize@url [0]{\catcode `\\12\catcode `\$12\catcode
  `\&12\catcode `\#12\catcode `\^12\catcode `\_12\catcode `\%12\relax}%
\providecommand \@@startlink[1]{}%
\providecommand \@@endlink[0]{}%
\providecommand \url  [0]{\begingroup\@sanitize@url \@url }%
\providecommand \@url [1]{\endgroup\@href {#1}{\urlprefix }}%
\providecommand \urlprefix  [0]{URL }%
\providecommand \Eprint [0]{\href }%
\providecommand \doibase [0]{https://doi.org/}%
\providecommand \selectlanguage [0]{\@gobble}%
\providecommand \bibinfo  [0]{\@secondoftwo}%
\providecommand \bibfield  [0]{\@secondoftwo}%
\providecommand \translation [1]{[#1]}%
\providecommand \BibitemOpen [0]{}%
\providecommand \bibitemStop [0]{}%
\providecommand \bibitemNoStop [0]{.\EOS\space}%
\providecommand \EOS [0]{\spacefactor3000\relax}%
\providecommand \BibitemShut  [1]{\csname bibitem#1\endcsname}%
\let\auto@bib@innerbib\@empty
\bibitem [{\citenamefont {Serafinelli}\ \emph {et~al.}(2022)\citenamefont
  {Serafinelli}, \citenamefont {Fantoni}, \citenamefont {Alegria},\ and\
  \citenamefont {Vieira}}]{serafinelli2022plasmonic}%
  \BibitemOpen
  \bibfield  {author} {\bibinfo {author} {\bibfnamefont {C.}~\bibnamefont
  {Serafinelli}}, \bibinfo {author} {\bibfnamefont {A.}~\bibnamefont
  {Fantoni}}, \bibinfo {author} {\bibfnamefont {E.~C.}\ \bibnamefont
  {Alegria}},\ and\ \bibinfo {author} {\bibfnamefont {M.}~\bibnamefont
  {Vieira}},\ }\bibfield  {title} {\bibinfo {title} {Plasmonic metal
  nanoparticles hybridized with {2D} nanomaterials for {SERS} detection: A
  review},\ }\href@noop {} {\bibfield  {journal} {\bibinfo  {journal}
  {Biosensors}\ }\textbf {\bibinfo {volume} {12}},\ \bibinfo {pages} {225}
  (\bibinfo {year} {2022})}\BibitemShut {NoStop}%
\bibitem [{\citenamefont {Chang}\ \emph {et~al.}(2007)\citenamefont {Chang},
  \citenamefont {S{\o}rensen}, \citenamefont {Demler},\ and\ \citenamefont
  {Lukin}}]{chang2007single}%
  \BibitemOpen
  \bibfield  {author} {\bibinfo {author} {\bibfnamefont {D.~E.}\ \bibnamefont
  {Chang}}, \bibinfo {author} {\bibfnamefont {A.~S.}\ \bibnamefont
  {S{\o}rensen}}, \bibinfo {author} {\bibfnamefont {E.~A.}\ \bibnamefont
  {Demler}},\ and\ \bibinfo {author} {\bibfnamefont {M.~D.}\ \bibnamefont
  {Lukin}},\ }\bibfield  {title} {\bibinfo {title} {A single-photon transistor
  using nanoscale surface plasmons},\ }\href@noop {} {\bibfield  {journal}
  {\bibinfo  {journal} {Nature Physics}\ }\textbf {\bibinfo {volume} {3}},\
  \bibinfo {pages} {807} (\bibinfo {year} {2007})}\BibitemShut {NoStop}%
\bibitem [{\citenamefont {Akimov}\ \emph {et~al.}(2007)\citenamefont {Akimov},
  \citenamefont {Mukherjee}, \citenamefont {Yu}, \citenamefont {Chang},
  \citenamefont {Zibrov}, \citenamefont {Hemmer}, \citenamefont {Park},\ and\
  \citenamefont {Lukin}}]{akimov2007generation}%
  \BibitemOpen
  \bibfield  {author} {\bibinfo {author} {\bibfnamefont {A.}~\bibnamefont
  {Akimov}}, \bibinfo {author} {\bibfnamefont {A.}~\bibnamefont {Mukherjee}},
  \bibinfo {author} {\bibfnamefont {C.}~\bibnamefont {Yu}}, \bibinfo {author}
  {\bibfnamefont {D.}~\bibnamefont {Chang}}, \bibinfo {author} {\bibfnamefont
  {A.}~\bibnamefont {Zibrov}}, \bibinfo {author} {\bibfnamefont
  {P.}~\bibnamefont {Hemmer}}, \bibinfo {author} {\bibfnamefont
  {H.}~\bibnamefont {Park}},\ and\ \bibinfo {author} {\bibfnamefont
  {M.}~\bibnamefont {Lukin}},\ }\bibfield  {title} {\bibinfo {title}
  {Generation of single optical plasmons in metallic nanowires coupled to
  quantum dots},\ }\href@noop {} {\bibfield  {journal} {\bibinfo  {journal}
  {Nature}\ }\textbf {\bibinfo {volume} {450}},\ \bibinfo {pages} {402}
  (\bibinfo {year} {2007})}\BibitemShut {NoStop}%
\bibitem [{\citenamefont {Purcell}(1995)}]{purcell1995spontaneous}%
  \BibitemOpen
  \bibfield  {author} {\bibinfo {author} {\bibfnamefont {E.~M.}\ \bibnamefont
  {Purcell}},\ }\bibfield  {title} {\bibinfo {title} {Spontaneous emission
  probabilities at radio frequencies},\ }in\ \href@noop {} {\emph {\bibinfo
  {booktitle} {Confined Electrons and Photons: New Physics and Applications}}}\
  (\bibinfo  {publisher} {Springer},\ \bibinfo {year} {1995})\ pp.\ \bibinfo
  {pages} {839--839}\BibitemShut {NoStop}%
\bibitem [{\citenamefont {Vahala}(2003)}]{vahala2003optical}%
  \BibitemOpen
  \bibfield  {author} {\bibinfo {author} {\bibfnamefont {K.~J.}\ \bibnamefont
  {Vahala}},\ }\bibfield  {title} {\bibinfo {title} {Optical microcavities},\
  }\href@noop {} {\bibfield  {journal} {\bibinfo  {journal} {Nature}\ }\textbf
  {\bibinfo {volume} {424}},\ \bibinfo {pages} {839} (\bibinfo {year}
  {2003})}\BibitemShut {NoStop}%
\bibitem [{\citenamefont {Lakowicz}(2005)}]{lakowicz2005radiative}%
  \BibitemOpen
  \bibfield  {author} {\bibinfo {author} {\bibfnamefont {J.~R.}\ \bibnamefont
  {Lakowicz}},\ }\bibfield  {title} {\bibinfo {title} {Radiative decay
  engineering 5: metal-enhanced fluorescence and plasmon emission},\
  }\href@noop {} {\bibfield  {journal} {\bibinfo  {journal} {Analytical
  Biochemistry}\ }\textbf {\bibinfo {volume} {337}},\ \bibinfo {pages} {171}
  (\bibinfo {year} {2005})}\BibitemShut {NoStop}%
\bibitem [{\citenamefont {Camposeo}\ \emph {et~al.}(2015)\citenamefont
  {Camposeo}, \citenamefont {Persano}, \citenamefont {Manco}, \citenamefont
  {Wang}, \citenamefont {Del~Carro}, \citenamefont {Zhang}, \citenamefont {Li},
  \citenamefont {Pisignano},\ and\ \citenamefont {Xia}}]{camposeo2015metal}%
  \BibitemOpen
  \bibfield  {author} {\bibinfo {author} {\bibfnamefont {A.}~\bibnamefont
  {Camposeo}}, \bibinfo {author} {\bibfnamefont {L.}~\bibnamefont {Persano}},
  \bibinfo {author} {\bibfnamefont {R.}~\bibnamefont {Manco}}, \bibinfo
  {author} {\bibfnamefont {Y.}~\bibnamefont {Wang}}, \bibinfo {author}
  {\bibfnamefont {P.}~\bibnamefont {Del~Carro}}, \bibinfo {author}
  {\bibfnamefont {C.}~\bibnamefont {Zhang}}, \bibinfo {author} {\bibfnamefont
  {Z.-Y.}\ \bibnamefont {Li}}, \bibinfo {author} {\bibfnamefont
  {D.}~\bibnamefont {Pisignano}},\ and\ \bibinfo {author} {\bibfnamefont
  {Y.}~\bibnamefont {Xia}},\ }\bibfield  {title} {\bibinfo {title}
  {Metal-enhanced near-infrared fluorescence by micropatterned gold
  nanocages},\ }\href@noop {} {\bibfield  {journal} {\bibinfo  {journal} {ACS
  Nano}\ }\textbf {\bibinfo {volume} {9}},\ \bibinfo {pages} {10047} (\bibinfo
  {year} {2015})}\BibitemShut {NoStop}%
\bibitem [{\citenamefont {Lu}\ \emph {et~al.}(2017)\citenamefont {Lu},
  \citenamefont {Sokhoyan}, \citenamefont {Cheng}, \citenamefont
  {Kafaie~Shirmanesh}, \citenamefont {Davoyan}, \citenamefont {Pala},
  \citenamefont {Thyagarajan},\ and\ \citenamefont
  {Atwater}}]{lu2017dynamically}%
  \BibitemOpen
  \bibfield  {author} {\bibinfo {author} {\bibfnamefont {Y.-J.}\ \bibnamefont
  {Lu}}, \bibinfo {author} {\bibfnamefont {R.}~\bibnamefont {Sokhoyan}},
  \bibinfo {author} {\bibfnamefont {W.-H.}\ \bibnamefont {Cheng}}, \bibinfo
  {author} {\bibfnamefont {G.}~\bibnamefont {Kafaie~Shirmanesh}}, \bibinfo
  {author} {\bibfnamefont {A.~R.}\ \bibnamefont {Davoyan}}, \bibinfo {author}
  {\bibfnamefont {R.~A.}\ \bibnamefont {Pala}}, \bibinfo {author}
  {\bibfnamefont {K.}~\bibnamefont {Thyagarajan}},\ and\ \bibinfo {author}
  {\bibfnamefont {H.~A.}\ \bibnamefont {Atwater}},\ }\bibfield  {title}
  {\bibinfo {title} {Dynamically controlled purcell enhancement of visible
  spontaneous emission in a gated plasmonic heterostructure},\ }\href@noop {}
  {\bibfield  {journal} {\bibinfo  {journal} {Nature Communications}\ }\textbf
  {\bibinfo {volume} {8}},\ \bibinfo {pages} {1631} (\bibinfo {year}
  {2017})}\BibitemShut {NoStop}%
\bibitem [{\citenamefont {Munnix}\ \emph {et~al.}(2009)\citenamefont {Munnix},
  \citenamefont {Lochmann}, \citenamefont {Bimberg},\ and\ \citenamefont
  {Haisler}}]{munnix2009modeling}%
  \BibitemOpen
  \bibfield  {author} {\bibinfo {author} {\bibfnamefont {M.~C.}\ \bibnamefont
  {Munnix}}, \bibinfo {author} {\bibfnamefont {A.}~\bibnamefont {Lochmann}},
  \bibinfo {author} {\bibfnamefont {D.}~\bibnamefont {Bimberg}},\ and\ \bibinfo
  {author} {\bibfnamefont {V.~A.}\ \bibnamefont {Haisler}},\ }\bibfield
  {title} {\bibinfo {title} {Modeling highly efficient rcled-type
  quantum-dot-based single photon emitters},\ }\href@noop {} {\bibfield
  {journal} {\bibinfo  {journal} {IEEE Journal of Quantum Electronics}\
  }\textbf {\bibinfo {volume} {45}},\ \bibinfo {pages} {1084} (\bibinfo {year}
  {2009})}\BibitemShut {NoStop}%
\bibitem [{\citenamefont {Bimberg}\ \emph {et~al.}(2009)\citenamefont
  {Bimberg}, \citenamefont {Stock}, \citenamefont {Lochmann}, \citenamefont
  {Schliwa}, \citenamefont {Tofflinger}, \citenamefont {Unrau}, \citenamefont
  {Munnix}, \citenamefont {Rodt}, \citenamefont {Haisler}, \citenamefont
  {Toropov} \emph {et~al.}}]{bimberg2009quantum}%
  \BibitemOpen
  \bibfield  {author} {\bibinfo {author} {\bibfnamefont {D.}~\bibnamefont
  {Bimberg}}, \bibinfo {author} {\bibfnamefont {E.}~\bibnamefont {Stock}},
  \bibinfo {author} {\bibfnamefont {A.}~\bibnamefont {Lochmann}}, \bibinfo
  {author} {\bibfnamefont {A.}~\bibnamefont {Schliwa}}, \bibinfo {author}
  {\bibfnamefont {J.~A.}\ \bibnamefont {Tofflinger}}, \bibinfo {author}
  {\bibfnamefont {W.}~\bibnamefont {Unrau}}, \bibinfo {author} {\bibfnamefont
  {M.}~\bibnamefont {Munnix}}, \bibinfo {author} {\bibfnamefont
  {S.}~\bibnamefont {Rodt}}, \bibinfo {author} {\bibfnamefont {V.~A.}\
  \bibnamefont {Haisler}}, \bibinfo {author} {\bibfnamefont {A.~I.}\
  \bibnamefont {Toropov}}, \emph {et~al.},\ }\bibfield  {title} {\bibinfo
  {title} {Quantum dots for single-and entangled-photon emitters},\ }\href@noop
  {} {\bibfield  {journal} {\bibinfo  {journal} {IEEE Photonics Journal}\
  }\textbf {\bibinfo {volume} {1}},\ \bibinfo {pages} {58} (\bibinfo {year}
  {2009})}\BibitemShut {NoStop}%
\bibitem [{\citenamefont {Canet-Ferrer}\ \emph {et~al.}(2012)\citenamefont
  {Canet-Ferrer}, \citenamefont {Mart{\'\i}nez}, \citenamefont {Prieto},
  \citenamefont {Al{\'e}n}, \citenamefont {Mu{\~n}oz-Matutano}, \citenamefont
  {Fuster}, \citenamefont {Gonz{\'a}lez}, \citenamefont {Dotor}, \citenamefont
  {Gonz{\'a}lez}, \citenamefont {Postigo} \emph {et~al.}}]{canet2012purcell}%
  \BibitemOpen
  \bibfield  {author} {\bibinfo {author} {\bibfnamefont {J.}~\bibnamefont
  {Canet-Ferrer}}, \bibinfo {author} {\bibfnamefont {L.~J.}\ \bibnamefont
  {Mart{\'\i}nez}}, \bibinfo {author} {\bibfnamefont {I.}~\bibnamefont
  {Prieto}}, \bibinfo {author} {\bibfnamefont {B.}~\bibnamefont {Al{\'e}n}},
  \bibinfo {author} {\bibfnamefont {G.}~\bibnamefont {Mu{\~n}oz-Matutano}},
  \bibinfo {author} {\bibfnamefont {D.}~\bibnamefont {Fuster}}, \bibinfo
  {author} {\bibfnamefont {Y.}~\bibnamefont {Gonz{\'a}lez}}, \bibinfo {author}
  {\bibfnamefont {M.~L.}\ \bibnamefont {Dotor}}, \bibinfo {author}
  {\bibfnamefont {L.}~\bibnamefont {Gonz{\'a}lez}}, \bibinfo {author}
  {\bibfnamefont {P.~A.}\ \bibnamefont {Postigo}}, \emph {et~al.},\ }\bibfield
  {title} {\bibinfo {title} {Purcell effect in photonic crystal microcavities
  embedding {InAs/InP} quantum wires},\ }\href@noop {} {\bibfield  {journal}
  {\bibinfo  {journal} {Optics Express}\ }\textbf {\bibinfo {volume} {20}},\
  \bibinfo {pages} {7901} (\bibinfo {year} {2012})}\BibitemShut {NoStop}%
\bibitem [{\citenamefont {Bennett}\ \emph {et~al.}(2005)\citenamefont
  {Bennett}, \citenamefont {Unitt}, \citenamefont {See}, \citenamefont
  {Shields}, \citenamefont {Atkinson}, \citenamefont {Cooper},\ and\
  \citenamefont {Ritchie}}]{bennett2005microcavity}%
  \BibitemOpen
  \bibfield  {author} {\bibinfo {author} {\bibfnamefont {A.}~\bibnamefont
  {Bennett}}, \bibinfo {author} {\bibfnamefont {D.}~\bibnamefont {Unitt}},
  \bibinfo {author} {\bibfnamefont {P.}~\bibnamefont {See}}, \bibinfo {author}
  {\bibfnamefont {A.}~\bibnamefont {Shields}}, \bibinfo {author} {\bibfnamefont
  {P.}~\bibnamefont {Atkinson}}, \bibinfo {author} {\bibfnamefont
  {K.}~\bibnamefont {Cooper}},\ and\ \bibinfo {author} {\bibfnamefont
  {D.}~\bibnamefont {Ritchie}},\ }\bibfield  {title} {\bibinfo {title}
  {Microcavity single-photon-emitting diode},\ }\href@noop {} {\bibfield
  {journal} {\bibinfo  {journal} {Applied Physics Letters}\ }\textbf {\bibinfo
  {volume} {86}} (\bibinfo {year} {2005})}\BibitemShut {NoStop}%
\bibitem [{\citenamefont {Moritake}\ \emph {et~al.}(2016)\citenamefont
  {Moritake}, \citenamefont {Kanamori},\ and\ \citenamefont
  {Hane}}]{moritake2016emission}%
  \BibitemOpen
  \bibfield  {author} {\bibinfo {author} {\bibfnamefont {Y.}~\bibnamefont
  {Moritake}}, \bibinfo {author} {\bibfnamefont {Y.}~\bibnamefont {Kanamori}},\
  and\ \bibinfo {author} {\bibfnamefont {K.}~\bibnamefont {Hane}},\ }\bibfield
  {title} {\bibinfo {title} {Emission wavelength tuning of fluorescence by fine
  structural control of optical metamaterials with fano resonance},\
  }\href@noop {} {\bibfield  {journal} {\bibinfo  {journal} {Scientific
  Reports}\ }\textbf {\bibinfo {volume} {6}},\ \bibinfo {pages} {33208}
  (\bibinfo {year} {2016})}\BibitemShut {NoStop}%
\bibitem [{\citenamefont {Steiner}\ \emph {et~al.}(2008)\citenamefont
  {Steiner}, \citenamefont {Failla}, \citenamefont {Hartschuh}, \citenamefont
  {Schleifenbaum}, \citenamefont {Stupperich},\ and\ \citenamefont
  {Meixner}}]{steiner2008controlling}%
  \BibitemOpen
  \bibfield  {author} {\bibinfo {author} {\bibfnamefont {M.}~\bibnamefont
  {Steiner}}, \bibinfo {author} {\bibfnamefont {A.~V.}\ \bibnamefont {Failla}},
  \bibinfo {author} {\bibfnamefont {A.}~\bibnamefont {Hartschuh}}, \bibinfo
  {author} {\bibfnamefont {F.}~\bibnamefont {Schleifenbaum}}, \bibinfo {author}
  {\bibfnamefont {C.}~\bibnamefont {Stupperich}},\ and\ \bibinfo {author}
  {\bibfnamefont {A.~J.}\ \bibnamefont {Meixner}},\ }\bibfield  {title}
  {\bibinfo {title} {Controlling molecular broadband-emission by optical
  confinement},\ }\href@noop {} {\bibfield  {journal} {\bibinfo  {journal} {New
  Journal of Physics}\ }\textbf {\bibinfo {volume} {10}},\ \bibinfo {pages}
  {123017} (\bibinfo {year} {2008})}\BibitemShut {NoStop}%
\bibitem [{\citenamefont {Chizhik}\ \emph {et~al.}(2009)\citenamefont
  {Chizhik}, \citenamefont {Schleifenbaum}, \citenamefont {Gutbrod},
  \citenamefont {Chizhik}, \citenamefont {Khoptyar}, \citenamefont {Meixner},\
  and\ \citenamefont {Enderlein}}]{chizhik2009tuning}%
  \BibitemOpen
  \bibfield  {author} {\bibinfo {author} {\bibfnamefont {A.}~\bibnamefont
  {Chizhik}}, \bibinfo {author} {\bibfnamefont {F.}~\bibnamefont
  {Schleifenbaum}}, \bibinfo {author} {\bibfnamefont {R.}~\bibnamefont
  {Gutbrod}}, \bibinfo {author} {\bibfnamefont {A.}~\bibnamefont {Chizhik}},
  \bibinfo {author} {\bibfnamefont {D.}~\bibnamefont {Khoptyar}}, \bibinfo
  {author} {\bibfnamefont {A.~J.}\ \bibnamefont {Meixner}},\ and\ \bibinfo
  {author} {\bibfnamefont {J.}~\bibnamefont {Enderlein}},\ }\bibfield  {title}
  {\bibinfo {title} {Tuning the fluorescence emission spectra of a single
  molecule with a variable optical subwavelength metal microcavity},\
  }\href@noop {} {\bibfield  {journal} {\bibinfo  {journal} {Physical Review
  Letters}\ }\textbf {\bibinfo {volume} {102}},\ \bibinfo {pages} {073002}
  (\bibinfo {year} {2009})}\BibitemShut {NoStop}%
\bibitem [{\citenamefont {Empedocles}\ and\ \citenamefont
  {Bawendi}(1997)}]{dots1997quantum}%
  \BibitemOpen
  \bibfield  {author} {\bibinfo {author} {\bibfnamefont {S.~A.}\ \bibnamefont
  {Empedocles}}\ and\ \bibinfo {author} {\bibfnamefont {M.~G.}\ \bibnamefont
  {Bawendi}},\ }\bibfield  {title} {\bibinfo {title} {Quantum-confined stark
  effect in single cdse nanocrystallite quantum dots},\ }\href@noop {}
  {\bibfield  {journal} {\bibinfo  {journal} {Science}\ }\textbf {\bibinfo
  {volume} {278}},\ \bibinfo {pages} {2114} (\bibinfo {year}
  {1997})}\BibitemShut {NoStop}%
\bibitem [{\citenamefont {Xu}\ \emph {et~al.}(2008)\citenamefont {Xu},
  \citenamefont {Andreev},\ and\ \citenamefont
  {Williams}}]{xu2008manipulating}%
  \BibitemOpen
  \bibfield  {author} {\bibinfo {author} {\bibfnamefont {X.}~\bibnamefont
  {Xu}}, \bibinfo {author} {\bibfnamefont {A.}~\bibnamefont {Andreev}},\ and\
  \bibinfo {author} {\bibfnamefont {D.~A.}\ \bibnamefont {Williams}},\
  }\bibfield  {title} {\bibinfo {title} {Manipulating quantum-confined stark
  shift in electroluminescence from quantum dots with side gates},\ }\href@noop
  {} {\bibfield  {journal} {\bibinfo  {journal} {New Journal of Physics}\
  }\textbf {\bibinfo {volume} {10}},\ \bibinfo {pages} {053036} (\bibinfo
  {year} {2008})}\BibitemShut {NoStop}%
\bibitem [{\citenamefont {Bozyigit}\ \emph {et~al.}(2013)\citenamefont
  {Bozyigit}, \citenamefont {Yarema},\ and\ \citenamefont
  {Wood}}]{bozyigit2013origins}%
  \BibitemOpen
  \bibfield  {author} {\bibinfo {author} {\bibfnamefont {D.}~\bibnamefont
  {Bozyigit}}, \bibinfo {author} {\bibfnamefont {O.}~\bibnamefont {Yarema}},\
  and\ \bibinfo {author} {\bibfnamefont {V.}~\bibnamefont {Wood}},\ }\bibfield
  {title} {\bibinfo {title} {Origins of low quantum efficiencies in quantum dot
  leds},\ }\href@noop {} {\bibfield  {journal} {\bibinfo  {journal} {Advanced
  Functional Materials}\ }\textbf {\bibinfo {volume} {23}},\ \bibinfo {pages}
  {3024} (\bibinfo {year} {2013})}\BibitemShut {NoStop}%
\bibitem [{\citenamefont {Salihoglu}\ \emph {et~al.}(2016)\citenamefont
  {Salihoglu}, \citenamefont {Kakenov}, \citenamefont {Balci}, \citenamefont
  {Balci},\ and\ \citenamefont {Kocabas}}]{salihoglu2016graphene}%
  \BibitemOpen
  \bibfield  {author} {\bibinfo {author} {\bibfnamefont {O.}~\bibnamefont
  {Salihoglu}}, \bibinfo {author} {\bibfnamefont {N.}~\bibnamefont {Kakenov}},
  \bibinfo {author} {\bibfnamefont {O.}~\bibnamefont {Balci}}, \bibinfo
  {author} {\bibfnamefont {S.}~\bibnamefont {Balci}},\ and\ \bibinfo {author}
  {\bibfnamefont {C.}~\bibnamefont {Kocabas}},\ }\bibfield  {title} {\bibinfo
  {title} {Graphene as a reversible and spectrally selective fluorescence
  quencher},\ }\href@noop {} {\bibfield  {journal} {\bibinfo  {journal}
  {Scientific Reports}\ }\textbf {\bibinfo {volume} {6}},\ \bibinfo {pages}
  {33911} (\bibinfo {year} {2016})}\BibitemShut {NoStop}%
\bibitem [{\citenamefont {Chen}\ \emph {et~al.}(2012)\citenamefont {Chen},
  \citenamefont {Gao}, \citenamefont {Zhang}, \citenamefont {Wu}, \citenamefont
  {Xiao},\ and\ \citenamefont {Jia}}]{chen2012electric}%
  \BibitemOpen
  \bibfield  {author} {\bibinfo {author} {\bibfnamefont {R.}~\bibnamefont
  {Chen}}, \bibinfo {author} {\bibfnamefont {Y.}~\bibnamefont {Gao}}, \bibinfo
  {author} {\bibfnamefont {G.}~\bibnamefont {Zhang}}, \bibinfo {author}
  {\bibfnamefont {R.}~\bibnamefont {Wu}}, \bibinfo {author} {\bibfnamefont
  {L.}~\bibnamefont {Xiao}},\ and\ \bibinfo {author} {\bibfnamefont
  {S.}~\bibnamefont {Jia}},\ }\bibfield  {title} {\bibinfo {title} {Electric
  field induced fluorescence modulation of single molecules in pmma based on
  electron transfer},\ }\href@noop {} {\bibfield  {journal} {\bibinfo
  {journal} {International Journal of Molecular Sciences}\ }\textbf {\bibinfo
  {volume} {13}},\ \bibinfo {pages} {11130} (\bibinfo {year}
  {2012})}\BibitemShut {NoStop}%
\bibitem [{\citenamefont {Wang}\ \emph {et~al.}(2013)\citenamefont {Wang},
  \citenamefont {Berda}, \citenamefont {Lu}, \citenamefont {Li}, \citenamefont
  {Wang},\ and\ \citenamefont {Chao}}]{wang2013tuning}%
  \BibitemOpen
  \bibfield  {author} {\bibinfo {author} {\bibfnamefont {S.}~\bibnamefont
  {Wang}}, \bibinfo {author} {\bibfnamefont {E.~B.}\ \bibnamefont {Berda}},
  \bibinfo {author} {\bibfnamefont {X.}~\bibnamefont {Lu}}, \bibinfo {author}
  {\bibfnamefont {X.}~\bibnamefont {Li}}, \bibinfo {author} {\bibfnamefont
  {C.}~\bibnamefont {Wang}},\ and\ \bibinfo {author} {\bibfnamefont
  {D.}~\bibnamefont {Chao}},\ }\bibfield  {title} {\bibinfo {title} {Tuning the
  fluorescent response of a novel electroactive polymer with multiple
  stimuli},\ }\href@noop {} {\bibfield  {journal} {\bibinfo  {journal}
  {Macromolecular Rapid Communications}\ }\textbf {\bibinfo {volume} {34}},\
  \bibinfo {pages} {1648} (\bibinfo {year} {2013})}\BibitemShut {NoStop}%
\bibitem [{\citenamefont {Xie}\ \emph {et~al.}(2022)\citenamefont {Xie},
  \citenamefont {Zhu}, \citenamefont {Li},\ and\ \citenamefont
  {Bulovi{\'c}}}]{xie2022voltage}%
  \BibitemOpen
  \bibfield  {author} {\bibinfo {author} {\bibfnamefont {S.}~\bibnamefont
  {Xie}}, \bibinfo {author} {\bibfnamefont {H.}~\bibnamefont {Zhu}}, \bibinfo
  {author} {\bibfnamefont {M.}~\bibnamefont {Li}},\ and\ \bibinfo {author}
  {\bibfnamefont {V.}~\bibnamefont {Bulovi{\'c}}},\ }\bibfield  {title}
  {\bibinfo {title} {Voltage-controlled reversible modulation of colloidal
  quantum dot thin film photoluminescence},\ }\href@noop {} {\bibfield
  {journal} {\bibinfo  {journal} {Applied Physics Letters}\ }\textbf {\bibinfo
  {volume} {120}} (\bibinfo {year} {2022})}\BibitemShut {NoStop}%
\bibitem [{\citenamefont {Leng}\ \emph {et~al.}(2018)\citenamefont {Leng},
  \citenamefont {Szychowski}, \citenamefont {Daniel},\ and\ \citenamefont
  {Pelton}}]{leng2018strong}%
  \BibitemOpen
  \bibfield  {author} {\bibinfo {author} {\bibfnamefont {H.}~\bibnamefont
  {Leng}}, \bibinfo {author} {\bibfnamefont {B.}~\bibnamefont {Szychowski}},
  \bibinfo {author} {\bibfnamefont {M.-C.}\ \bibnamefont {Daniel}},\ and\
  \bibinfo {author} {\bibfnamefont {M.}~\bibnamefont {Pelton}},\ }\bibfield
  {title} {\bibinfo {title} {Strong coupling and induced transparency at room
  temperature with single quantum dots and gap plasmons},\ }\href@noop {}
  {\bibfield  {journal} {\bibinfo  {journal} {Nature Communications}\ }\textbf
  {\bibinfo {volume} {9}},\ \bibinfo {pages} {4012} (\bibinfo {year}
  {2018})}\BibitemShut {NoStop}%
\bibitem [{\citenamefont {Limonov}\ \emph {et~al.}(2017)\citenamefont
  {Limonov}, \citenamefont {Rybin}, \citenamefont {Poddubny},\ and\
  \citenamefont {Kivshar}}]{limonov2017fano}%
  \BibitemOpen
  \bibfield  {author} {\bibinfo {author} {\bibfnamefont {M.~F.}\ \bibnamefont
  {Limonov}}, \bibinfo {author} {\bibfnamefont {M.~V.}\ \bibnamefont {Rybin}},
  \bibinfo {author} {\bibfnamefont {A.~N.}\ \bibnamefont {Poddubny}},\ and\
  \bibinfo {author} {\bibfnamefont {Y.~S.}\ \bibnamefont {Kivshar}},\
  }\bibfield  {title} {\bibinfo {title} {Fano resonances in photonics},\
  }\href@noop {} {\bibfield  {journal} {\bibinfo  {journal} {Nature Photonics}\
  }\textbf {\bibinfo {volume} {11}},\ \bibinfo {pages} {543} (\bibinfo {year}
  {2017})}\BibitemShut {NoStop}%
\bibitem [{\citenamefont {Tasgin}\ \emph {et~al.}(2018)\citenamefont {Tasgin},
  \citenamefont {Bek},\ and\ \citenamefont {Postac{\i}}}]{tacsgin2018fano}%
  \BibitemOpen
  \bibfield  {author} {\bibinfo {author} {\bibfnamefont {M.~E.}\ \bibnamefont
  {Tasgin}}, \bibinfo {author} {\bibfnamefont {A.}~\bibnamefont {Bek}},\ and\
  \bibinfo {author} {\bibfnamefont {S.}~\bibnamefont {Postac{\i}}},\ }\bibfield
   {title} {\bibinfo {title} {Fano resonances in the linear and nonlinear
  plasmonic response},\ }\href
  {https://link.springer.com/chapter/10.1007/978-3-319-99731-5_1} {\bibfield
  {journal} {\bibinfo  {journal} {Fano Resonances in Optics and Microwaves:
  Physics and Applications}\ }\textbf {\bibinfo {volume} {219}} (\bibinfo
  {year} {2018})}\BibitemShut {NoStop}%
\bibitem [{\citenamefont {Murata}\ \emph {et~al.}(2011)\citenamefont {Murata},
  \citenamefont {Yasutake}, \citenamefont {Nittoh}, \citenamefont {Fukatsu},\
  and\ \citenamefont {Miki}}]{murata2011high}%
  \BibitemOpen
  \bibfield  {author} {\bibinfo {author} {\bibfnamefont {K.}~\bibnamefont
  {Murata}}, \bibinfo {author} {\bibfnamefont {Y.}~\bibnamefont {Yasutake}},
  \bibinfo {author} {\bibfnamefont {K.-i.}\ \bibnamefont {Nittoh}}, \bibinfo
  {author} {\bibfnamefont {S.}~\bibnamefont {Fukatsu}},\ and\ \bibinfo {author}
  {\bibfnamefont {K.}~\bibnamefont {Miki}},\ }\bibfield  {title} {\bibinfo
  {title} {High-density g-centers, light-emitting point defects in silicon
  crystal},\ }\href@noop {} {\bibfield  {journal} {\bibinfo  {journal} {AIP
  Advances}\ }\textbf {\bibinfo {volume} {1}} (\bibinfo {year}
  {2011})}\BibitemShut {NoStop}%
\bibitem [{\citenamefont {Rotem}\ \emph {et~al.}(2007)\citenamefont {Rotem},
  \citenamefont {Shainline},\ and\ \citenamefont {Xu}}]{rotem2007enhanced}%
  \BibitemOpen
  \bibfield  {author} {\bibinfo {author} {\bibfnamefont {E.}~\bibnamefont
  {Rotem}}, \bibinfo {author} {\bibfnamefont {J.~M.}\ \bibnamefont
  {Shainline}},\ and\ \bibinfo {author} {\bibfnamefont {J.~M.}\ \bibnamefont
  {Xu}},\ }\bibfield  {title} {\bibinfo {title} {Enhanced photoluminescence
  from nanopatterned carbon-rich silicon grown by solid-phase epitaxy},\
  }\href@noop {} {\bibfield  {journal} {\bibinfo  {journal} {Applied Physics
  Letters}\ }\textbf {\bibinfo {volume} {91}} (\bibinfo {year}
  {2007})}\BibitemShut {NoStop}%
\bibitem [{\citenamefont {Ngandeu~Ngambou}\ \emph {et~al.}(2024)\citenamefont
  {Ngandeu~Ngambou}, \citenamefont {Perrin}, \citenamefont {Balasa},
  \citenamefont {Tiranov}, \citenamefont {Brinza}, \citenamefont {Bénédic},
  \citenamefont {Renaud}, \citenamefont {Reveillard}, \citenamefont {Silvent},
  \citenamefont {Goldner}, \citenamefont {Achard},\ and\ \citenamefont
  {Tallaire}}]{ngandeu2024hot}%
  \BibitemOpen
  \bibfield  {author} {\bibinfo {author} {\bibfnamefont {M.~W.}\ \bibnamefont
  {Ngandeu~Ngambou}}, \bibinfo {author} {\bibfnamefont {P.}~\bibnamefont
  {Perrin}}, \bibinfo {author} {\bibfnamefont {I.}~\bibnamefont {Balasa}},
  \bibinfo {author} {\bibfnamefont {A.}~\bibnamefont {Tiranov}}, \bibinfo
  {author} {\bibfnamefont {O.}~\bibnamefont {Brinza}}, \bibinfo {author}
  {\bibfnamefont {F.}~\bibnamefont {Bénédic}}, \bibinfo {author}
  {\bibfnamefont {J.}~\bibnamefont {Renaud}}, \bibinfo {author} {\bibfnamefont
  {M.}~\bibnamefont {Reveillard}}, \bibinfo {author} {\bibfnamefont
  {J.}~\bibnamefont {Silvent}}, \bibinfo {author} {\bibfnamefont
  {P.}~\bibnamefont {Goldner}}, \bibinfo {author} {\bibfnamefont
  {J.}~\bibnamefont {Achard}},\ and\ \bibinfo {author} {\bibfnamefont
  {A.}~\bibnamefont {Tallaire}},\ }\bibfield  {title} {\bibinfo {title} {Hot
  ion implantation to create dense {NV} center ensembles in diamond},\ }\href
  {https://doi.org/10.1063/5.0196719} {\bibfield  {journal} {\bibinfo
  {journal} {Applied Physics Letters}\ }\textbf {\bibinfo {volume} {124}},\
  \bibinfo {pages} {134002} (\bibinfo {year} {2024})}\BibitemShut {NoStop}%
\bibitem [{Note1()}]{Note1}%
  \BibitemOpen
  \bibinfo {note} {We refer the frequencies by the corresponding free-space
  wavelength, governed by $\omega =c/\lambda $.}\BibitemShut {Stop}%
\bibitem [{\citenamefont {Ansys/Lumerical}()}]{Ansys2024}%
  \BibitemOpen
  \bibfield  {author} {\bibinfo {author} {\bibnamefont {Ansys/Lumerical}},\
  }\href@noop {} {\bibinfo {title} {Fluorescence enhancement}},\ \bibinfo
  {howpublished} {\url{https: //optics.ansys.com /hc/en-us/articles/
  360042161033- Fluorescence-enhancement}}\BibitemShut {NoStop}%
\bibitem [{\citenamefont {Wu}\ \emph {et~al.}(2010)\citenamefont {Wu},
  \citenamefont {Gray},\ and\ \citenamefont {Pelton}}]{wu2010quantum}%
  \BibitemOpen
  \bibfield  {author} {\bibinfo {author} {\bibfnamefont {X.}~\bibnamefont
  {Wu}}, \bibinfo {author} {\bibfnamefont {S.~K.}\ \bibnamefont {Gray}},\ and\
  \bibinfo {author} {\bibfnamefont {M.}~\bibnamefont {Pelton}},\ }\bibfield
  {title} {\bibinfo {title} {Quantum-dot-induced transparency in a nanoscale
  plasmonic resonator},\ }\href@noop {} {\bibfield  {journal} {\bibinfo
  {journal} {Optics Express}\ }\textbf {\bibinfo {volume} {18}},\ \bibinfo
  {pages} {23633} (\bibinfo {year} {2010})}\BibitemShut {NoStop}%
\bibitem [{\citenamefont {Fleischhauer}\ \emph {et~al.}(2005)\citenamefont
  {Fleischhauer}, \citenamefont {Imamoglu},\ and\ \citenamefont
  {Marangos}}]{fleischhauer2005electromagnetically}%
  \BibitemOpen
  \bibfield  {author} {\bibinfo {author} {\bibfnamefont {M.}~\bibnamefont
  {Fleischhauer}}, \bibinfo {author} {\bibfnamefont {A.}~\bibnamefont
  {Imamoglu}},\ and\ \bibinfo {author} {\bibfnamefont {J.~P.}\ \bibnamefont
  {Marangos}},\ }\bibfield  {title} {\bibinfo {title} {Electromagnetically
  induced transparency: Optics in coherent media},\ }\href@noop {} {\bibfield
  {journal} {\bibinfo  {journal} {Reviews of Modern Physics}\ }\textbf
  {\bibinfo {volume} {77}},\ \bibinfo {pages} {633} (\bibinfo {year}
  {2005})}\BibitemShut {NoStop}%
\bibitem [{\citenamefont {Shah}\ \emph {et~al.}(2013)\citenamefont {Shah},
  \citenamefont {Scherer}, \citenamefont {Pelton},\ and\ \citenamefont
  {Gray}}]{shah2013ultrafast}%
  \BibitemOpen
  \bibfield  {author} {\bibinfo {author} {\bibfnamefont {R.~A.}\ \bibnamefont
  {Shah}}, \bibinfo {author} {\bibfnamefont {N.~F.}\ \bibnamefont {Scherer}},
  \bibinfo {author} {\bibfnamefont {M.}~\bibnamefont {Pelton}},\ and\ \bibinfo
  {author} {\bibfnamefont {S.~K.}\ \bibnamefont {Gray}},\ }\bibfield  {title}
  {\bibinfo {title} {Ultrafast reversal of a fano resonance in a
  plasmon-exciton system},\ }\href@noop {} {\bibfield  {journal} {\bibinfo
  {journal} {Physical Review B—Condensed Matter and Materials Physics}\
  }\textbf {\bibinfo {volume} {88}},\ \bibinfo {pages} {075411} (\bibinfo
  {year} {2013})}\BibitemShut {NoStop}%
\bibitem [{\citenamefont {Garrido~Alzar}\ \emph {et~al.}(2002)\citenamefont
  {Garrido~Alzar}, \citenamefont {Martinez},\ and\ \citenamefont
  {Nussenzveig}}]{alzar2001classical}%
  \BibitemOpen
  \bibfield  {author} {\bibinfo {author} {\bibfnamefont {C.~L.}\ \bibnamefont
  {Garrido~Alzar}}, \bibinfo {author} {\bibfnamefont {M.~A.~G.}\ \bibnamefont
  {Martinez}},\ and\ \bibinfo {author} {\bibfnamefont {P.}~\bibnamefont
  {Nussenzveig}},\ }\bibfield  {title} {\bibinfo {title} {Classical analog of
  electromagnetically induced transparency},\ }\href
  {https://doi.org/10.1119/1.1412644} {\bibfield  {journal} {\bibinfo
  {journal} {American Journal of Physics}\ }\textbf {\bibinfo {volume} {70}},\
  \bibinfo {pages} {37} (\bibinfo {year} {2002})}\BibitemShut {NoStop}%
\bibitem [{SM()}]{SM}%
  \BibitemOpen
  \href@noop {} {}\bibinfo {note} {See Supplemental Material at
  "\url{https://}"}\BibitemShut {NoStop}%
\bibitem [{\citenamefont {Larocque}\ \emph {et~al.}(2024)\citenamefont
  {Larocque}, \citenamefont {Buyukkaya}, \citenamefont {Errando-Herranz},
  \citenamefont {Papon}, \citenamefont {Harper}, \citenamefont {Tao},
  \citenamefont {Carolan}, \citenamefont {Lee}, \citenamefont {Richardson},
  \citenamefont {Leake} \emph {et~al.}}]{larocque2024tunable}%
  \BibitemOpen
  \bibfield  {author} {\bibinfo {author} {\bibfnamefont {H.}~\bibnamefont
  {Larocque}}, \bibinfo {author} {\bibfnamefont {M.~A.}\ \bibnamefont
  {Buyukkaya}}, \bibinfo {author} {\bibfnamefont {C.}~\bibnamefont
  {Errando-Herranz}}, \bibinfo {author} {\bibfnamefont {C.}~\bibnamefont
  {Papon}}, \bibinfo {author} {\bibfnamefont {S.}~\bibnamefont {Harper}},
  \bibinfo {author} {\bibfnamefont {M.}~\bibnamefont {Tao}}, \bibinfo {author}
  {\bibfnamefont {J.}~\bibnamefont {Carolan}}, \bibinfo {author} {\bibfnamefont
  {C.-M.}\ \bibnamefont {Lee}}, \bibinfo {author} {\bibfnamefont {C.~J.}\
  \bibnamefont {Richardson}}, \bibinfo {author} {\bibfnamefont {G.~L.}\
  \bibnamefont {Leake}}, \emph {et~al.},\ }\bibfield  {title} {\bibinfo {title}
  {Tunable quantum emitters on large-scale foundry silicon photonics},\
  }\href@noop {} {\bibfield  {journal} {\bibinfo  {journal} {Nature
  Communications}\ }\textbf {\bibinfo {volume} {15}},\ \bibinfo {pages} {5781}
  (\bibinfo {year} {2024})}\BibitemShut {NoStop}%
\bibitem [{\citenamefont {Miller}\ \emph {et~al.}(1985)\citenamefont {Miller},
  \citenamefont {Chemla}, \citenamefont {Damen}, \citenamefont {Gossard},
  \citenamefont {Wiegmann}, \citenamefont {Wood},\ and\ \citenamefont
  {Burrus}}]{miller1985electric}%
  \BibitemOpen
  \bibfield  {author} {\bibinfo {author} {\bibfnamefont {D.~A.}\ \bibnamefont
  {Miller}}, \bibinfo {author} {\bibfnamefont {D.}~\bibnamefont {Chemla}},
  \bibinfo {author} {\bibfnamefont {T.}~\bibnamefont {Damen}}, \bibinfo
  {author} {\bibfnamefont {A.}~\bibnamefont {Gossard}}, \bibinfo {author}
  {\bibfnamefont {W.}~\bibnamefont {Wiegmann}}, \bibinfo {author}
  {\bibfnamefont {T.}~\bibnamefont {Wood}},\ and\ \bibinfo {author}
  {\bibfnamefont {C.}~\bibnamefont {Burrus}},\ }\bibfield  {title} {\bibinfo
  {title} {Electric field dependence of optical absorption near the band gap of
  quantum-well structures},\ }\href@noop {} {\bibfield  {journal} {\bibinfo
  {journal} {Physical Review B}\ }\textbf {\bibinfo {volume} {32}},\ \bibinfo
  {pages} {1043} (\bibinfo {year} {1985})}\BibitemShut {NoStop}%
\bibitem [{\citenamefont {Rajput}\ and\ \citenamefont
  {Shishodia}(2020)}]{rajput2020forster}%
  \BibitemOpen
  \bibfield  {author} {\bibinfo {author} {\bibfnamefont {P.}~\bibnamefont
  {Rajput}}\ and\ \bibinfo {author} {\bibfnamefont {M.~S.}\ \bibnamefont
  {Shishodia}},\ }\bibfield  {title} {\bibinfo {title} {F{\"o}rster resonance
  energy transfer and molecular fluorescence near gain assisted refractory
  nitrides based plasmonic core-shell nanoparticle},\ }\href@noop {} {\bibfield
   {journal} {\bibinfo  {journal} {Plasmonics}\ }\textbf {\bibinfo {volume}
  {15}},\ \bibinfo {pages} {2081} (\bibinfo {year} {2020})}\BibitemShut
  {NoStop}%
\bibitem [{\citenamefont {Issah}\ \emph {et~al.}(2023)\citenamefont {Issah},
  \citenamefont {Pietila}, \citenamefont {Kujala}, \citenamefont {Koivurova},
  \citenamefont {Caglayan},\ and\ \citenamefont
  {Ornigotti}}]{issah2023epsilon}%
  \BibitemOpen
  \bibfield  {author} {\bibinfo {author} {\bibfnamefont {I.}~\bibnamefont
  {Issah}}, \bibinfo {author} {\bibfnamefont {J.}~\bibnamefont {Pietila}},
  \bibinfo {author} {\bibfnamefont {T.}~\bibnamefont {Kujala}}, \bibinfo
  {author} {\bibfnamefont {M.}~\bibnamefont {Koivurova}}, \bibinfo {author}
  {\bibfnamefont {H.}~\bibnamefont {Caglayan}},\ and\ \bibinfo {author}
  {\bibfnamefont {M.}~\bibnamefont {Ornigotti}},\ }\bibfield  {title} {\bibinfo
  {title} {Epsilon-near-zero nanoparticles},\ }\href@noop {} {\bibfield
  {journal} {\bibinfo  {journal} {Physical Review A}\ }\textbf {\bibinfo
  {volume} {107}},\ \bibinfo {pages} {023501} (\bibinfo {year}
  {2023})}\BibitemShut {NoStop}%
\bibitem [{\citenamefont {Y{\"u}cel}\ \emph {et~al.}(2020)\citenamefont
  {Y{\"u}cel}, \citenamefont {Ate{\c{s}}},\ and\ \citenamefont
  {Bek}}]{yucel2020single}%
  \BibitemOpen
  \bibfield  {author} {\bibinfo {author} {\bibfnamefont {O.}~\bibnamefont
  {Y{\"u}cel}}, \bibinfo {author} {\bibfnamefont {S.}~\bibnamefont
  {Ate{\c{s}}}},\ and\ \bibinfo {author} {\bibfnamefont {A.}~\bibnamefont
  {Bek}},\ }\bibfield  {title} {\bibinfo {title} {Single-photon nanoantenna
  with in situ fabrication of plasmonic {Ag} nanoparticle at an {hBN} defect
  center},\ }\href@noop {} {\bibfield  {journal} {\bibinfo  {journal} {arXiv
  preprint arXiv:2003.13824}\ } (\bibinfo {year} {2020})}\BibitemShut {NoStop}%
\bibitem [{\citenamefont {Liu}\ \emph {et~al.}(2017)\citenamefont {Liu},
  \citenamefont {Li},\ and\ \citenamefont {Xiao}}]{liu2017electromagnetically}%
  \BibitemOpen
  \bibfield  {author} {\bibinfo {author} {\bibfnamefont {Y.-C.}\ \bibnamefont
  {Liu}}, \bibinfo {author} {\bibfnamefont {B.-B.}\ \bibnamefont {Li}},\ and\
  \bibinfo {author} {\bibfnamefont {Y.-F.}\ \bibnamefont {Xiao}},\ }\bibfield
  {title} {\bibinfo {title} {Electromagnetically induced transparency in
  optical microcavities},\ }\href@noop {} {\bibfield  {journal} {\bibinfo
  {journal} {Nanophotonics}\ }\textbf {\bibinfo {volume} {6}},\ \bibinfo
  {pages} {789} (\bibinfo {year} {2017})}\BibitemShut {NoStop}%
\bibitem [{\citenamefont {Johnson}\ and\ \citenamefont
  {Christy}(1972)}]{johnson1972optical}%
  \BibitemOpen
  \bibfield  {author} {\bibinfo {author} {\bibfnamefont {P.~B.}\ \bibnamefont
  {Johnson}}\ and\ \bibinfo {author} {\bibfnamefont {R.-W.}\ \bibnamefont
  {Christy}},\ }\bibfield  {title} {\bibinfo {title} {Optical constants of the
  noble metals},\ }\href@noop {} {\bibfield  {journal} {\bibinfo  {journal}
  {Physical Review B}\ }\textbf {\bibinfo {volume} {6}},\ \bibinfo {pages}
  {4370} (\bibinfo {year} {1972})}\BibitemShut {NoStop}%
\bibitem [{\citenamefont {Palik}(1998)}]{palik1998handbook}%
  \BibitemOpen
  \bibfield  {author} {\bibinfo {author} {\bibfnamefont {E.~D.}\ \bibnamefont
  {Palik}},\ }\href@noop {} {\emph {\bibinfo {title} {Handbook of optical
  constants of solids}}},\ Vol.~\bibinfo {volume} {3}\ (\bibinfo  {publisher}
  {Academic press},\ \bibinfo {year} {1998})\BibitemShut {NoStop}%
\bibitem [{\citenamefont {Shibata}\ \emph {et~al.}(2013)\citenamefont
  {Shibata}, \citenamefont {Yuan}, \citenamefont {Iwasa},\ and\ \citenamefont
  {Hirakawa}}]{shibata2013large}%
  \BibitemOpen
  \bibfield  {author} {\bibinfo {author} {\bibfnamefont {K.}~\bibnamefont
  {Shibata}}, \bibinfo {author} {\bibfnamefont {H.}~\bibnamefont {Yuan}},
  \bibinfo {author} {\bibfnamefont {Y.}~\bibnamefont {Iwasa}},\ and\ \bibinfo
  {author} {\bibfnamefont {K.}~\bibnamefont {Hirakawa}},\ }\bibfield  {title}
  {\bibinfo {title} {Large modulation of zero-dimensional electronic states in
  quantum dots by electric-double-layer gating},\ }\href@noop {} {\bibfield
  {journal} {\bibinfo  {journal} {Nature Communications}\ }\textbf {\bibinfo
  {volume} {4}},\ \bibinfo {pages} {2664} (\bibinfo {year} {2013})}\BibitemShut
  {NoStop}%
\bibitem [{\citenamefont {Tasgin}(2017)}]{tasgin2017many}%
  \BibitemOpen
  \bibfield  {author} {\bibinfo {author} {\bibfnamefont {M.~E.}\ \bibnamefont
  {Tasgin}},\ }\bibfield  {title} {\bibinfo {title} {Many-particle entanglement
  criterion for superradiantlike states},\ }\href@noop {} {\bibfield  {journal}
  {\bibinfo  {journal} {Physical Review Letters}\ }\textbf {\bibinfo {volume}
  {119}},\ \bibinfo {pages} {033601} (\bibinfo {year} {2017})}\BibitemShut
  {NoStop}%
\bibitem [{\citenamefont {Alicki}\ and\ \citenamefont
  {Fannes}(2013)}]{alicki2013entanglement}%
  \BibitemOpen
  \bibfield  {author} {\bibinfo {author} {\bibfnamefont {R.}~\bibnamefont
  {Alicki}}\ and\ \bibinfo {author} {\bibfnamefont {M.}~\bibnamefont
  {Fannes}},\ }\bibfield  {title} {\bibinfo {title} {Entanglement boost for
  extractable work from ensembles of quantum batteries},\ }\href@noop {}
  {\bibfield  {journal} {\bibinfo  {journal} {Physical Review E—Statistical,
  Nonlinear, and Soft Matter Physics}\ }\textbf {\bibinfo {volume} {87}},\
  \bibinfo {pages} {042123} (\bibinfo {year} {2013})}\BibitemShut {NoStop}%
\bibitem [{\citenamefont {Ferraro}\ \emph {et~al.}(2018)\citenamefont
  {Ferraro}, \citenamefont {Campisi}, \citenamefont {Andolina}, \citenamefont
  {Pellegrini},\ and\ \citenamefont {Polini}}]{ferraro2018high}%
  \BibitemOpen
  \bibfield  {author} {\bibinfo {author} {\bibfnamefont {D.}~\bibnamefont
  {Ferraro}}, \bibinfo {author} {\bibfnamefont {M.}~\bibnamefont {Campisi}},
  \bibinfo {author} {\bibfnamefont {G.~M.}\ \bibnamefont {Andolina}}, \bibinfo
  {author} {\bibfnamefont {V.}~\bibnamefont {Pellegrini}},\ and\ \bibinfo
  {author} {\bibfnamefont {M.}~\bibnamefont {Polini}},\ }\bibfield  {title}
  {\bibinfo {title} {High-power collective charging of a solid-state quantum
  battery},\ }\href@noop {} {\bibfield  {journal} {\bibinfo  {journal}
  {Physical Review Letters}\ }\textbf {\bibinfo {volume} {120}},\ \bibinfo
  {pages} {117702} (\bibinfo {year} {2018})}\BibitemShut {NoStop}%
\bibitem [{\citenamefont {Bluvstein}\ \emph {et~al.}(2024)\citenamefont
  {Bluvstein}, \citenamefont {Evered}, \citenamefont {Geim}, \citenamefont
  {Li}, \citenamefont {Zhou}, \citenamefont {Manovitz}, \citenamefont {Ebadi},
  \citenamefont {Cain}, \citenamefont {Kalinowski}, \citenamefont {Hangleiter}
  \emph {et~al.}}]{bluvstein2024logical}%
  \BibitemOpen
  \bibfield  {author} {\bibinfo {author} {\bibfnamefont {D.}~\bibnamefont
  {Bluvstein}}, \bibinfo {author} {\bibfnamefont {S.~J.}\ \bibnamefont
  {Evered}}, \bibinfo {author} {\bibfnamefont {A.~A.}\ \bibnamefont {Geim}},
  \bibinfo {author} {\bibfnamefont {S.~H.}\ \bibnamefont {Li}}, \bibinfo
  {author} {\bibfnamefont {H.}~\bibnamefont {Zhou}}, \bibinfo {author}
  {\bibfnamefont {T.}~\bibnamefont {Manovitz}}, \bibinfo {author}
  {\bibfnamefont {S.}~\bibnamefont {Ebadi}}, \bibinfo {author} {\bibfnamefont
  {M.}~\bibnamefont {Cain}}, \bibinfo {author} {\bibfnamefont {M.}~\bibnamefont
  {Kalinowski}}, \bibinfo {author} {\bibfnamefont {D.}~\bibnamefont
  {Hangleiter}}, \emph {et~al.},\ }\bibfield  {title} {\bibinfo {title}
  {Logical quantum processor based on reconfigurable atom arrays},\ }\href@noop
  {} {\bibfield  {journal} {\bibinfo  {journal} {Nature}\ }\textbf {\bibinfo
  {volume} {626}},\ \bibinfo {pages} {58} (\bibinfo {year} {2024})}\BibitemShut
  {NoStop}%
\bibitem [{\citenamefont {Lodahl}\ \emph {et~al.}(2004)\citenamefont {Lodahl},
  \citenamefont {Floris~van Driel}, \citenamefont {Nikolaev}, \citenamefont
  {Irman}, \citenamefont {Overgaag}, \citenamefont {Vanmaekelbergh},\ and\
  \citenamefont {Vos}}]{lodahl2004controlling}%
  \BibitemOpen
  \bibfield  {author} {\bibinfo {author} {\bibfnamefont {P.}~\bibnamefont
  {Lodahl}}, \bibinfo {author} {\bibfnamefont {A.}~\bibnamefont {Floris~van
  Driel}}, \bibinfo {author} {\bibfnamefont {I.~S.}\ \bibnamefont {Nikolaev}},
  \bibinfo {author} {\bibfnamefont {A.}~\bibnamefont {Irman}}, \bibinfo
  {author} {\bibfnamefont {K.}~\bibnamefont {Overgaag}}, \bibinfo {author}
  {\bibfnamefont {D.}~\bibnamefont {Vanmaekelbergh}},\ and\ \bibinfo {author}
  {\bibfnamefont {W.~L.}\ \bibnamefont {Vos}},\ }\bibfield  {title} {\bibinfo
  {title} {Controlling the dynamics of spontaneous emission from quantum dots
  by photonic crystals},\ }\href@noop {} {\bibfield  {journal} {\bibinfo
  {journal} {Nature}\ }\textbf {\bibinfo {volume} {430}},\ \bibinfo {pages}
  {654} (\bibinfo {year} {2004})}\BibitemShut {NoStop}%
\bibitem [{\citenamefont {Yuan}\ \emph {et~al.}(2002)\citenamefont {Yuan},
  \citenamefont {Kardynal}, \citenamefont {Stevenson}, \citenamefont {Shields},
  \citenamefont {Lobo}, \citenamefont {Cooper}, \citenamefont {Beattie},
  \citenamefont {Ritchie},\ and\ \citenamefont
  {Pepper}}]{yuan2002electrically}%
  \BibitemOpen
  \bibfield  {author} {\bibinfo {author} {\bibfnamefont {Z.}~\bibnamefont
  {Yuan}}, \bibinfo {author} {\bibfnamefont {B.~E.}\ \bibnamefont {Kardynal}},
  \bibinfo {author} {\bibfnamefont {R.~M.}\ \bibnamefont {Stevenson}}, \bibinfo
  {author} {\bibfnamefont {A.~J.}\ \bibnamefont {Shields}}, \bibinfo {author}
  {\bibfnamefont {C.~J.}\ \bibnamefont {Lobo}}, \bibinfo {author}
  {\bibfnamefont {K.}~\bibnamefont {Cooper}}, \bibinfo {author} {\bibfnamefont
  {N.~S.}\ \bibnamefont {Beattie}}, \bibinfo {author} {\bibfnamefont {D.~A.}\
  \bibnamefont {Ritchie}},\ and\ \bibinfo {author} {\bibfnamefont
  {M.}~\bibnamefont {Pepper}},\ }\bibfield  {title} {\bibinfo {title}
  {Electrically driven single-photon source},\ }\href@noop {} {\bibfield
  {journal} {\bibinfo  {journal} {science}\ }\textbf {\bibinfo {volume}
  {295}},\ \bibinfo {pages} {102} (\bibinfo {year} {2002})}\BibitemShut
  {NoStop}%
\bibitem [{\citenamefont {St{\"u}rner}\ \emph {et~al.}(2021)\citenamefont
  {St{\"u}rner}, \citenamefont {Brenneis}, \citenamefont {Buck}, \citenamefont
  {Kassel}, \citenamefont {R{\"o}lver}, \citenamefont {Fuchs}, \citenamefont
  {Savitsky}, \citenamefont {Suter}, \citenamefont {Grimmel}, \citenamefont
  {Hengesbach} \emph {et~al.}}]{sturner2021integrated}%
  \BibitemOpen
  \bibfield  {author} {\bibinfo {author} {\bibfnamefont {F.~M.}\ \bibnamefont
  {St{\"u}rner}}, \bibinfo {author} {\bibfnamefont {A.}~\bibnamefont
  {Brenneis}}, \bibinfo {author} {\bibfnamefont {T.}~\bibnamefont {Buck}},
  \bibinfo {author} {\bibfnamefont {J.}~\bibnamefont {Kassel}}, \bibinfo
  {author} {\bibfnamefont {R.}~\bibnamefont {R{\"o}lver}}, \bibinfo {author}
  {\bibfnamefont {T.}~\bibnamefont {Fuchs}}, \bibinfo {author} {\bibfnamefont
  {A.}~\bibnamefont {Savitsky}}, \bibinfo {author} {\bibfnamefont
  {D.}~\bibnamefont {Suter}}, \bibinfo {author} {\bibfnamefont
  {J.}~\bibnamefont {Grimmel}}, \bibinfo {author} {\bibfnamefont
  {S.}~\bibnamefont {Hengesbach}}, \emph {et~al.},\ }\bibfield  {title}
  {\bibinfo {title} {Integrated and portable magnetometer based on
  nitrogen-vacancy ensembles in diamond},\ }\href@noop {} {\bibfield  {journal}
  {\bibinfo  {journal} {Advanced Quantum Technologies}\ }\textbf {\bibinfo
  {volume} {4}},\ \bibinfo {pages} {2000111} (\bibinfo {year}
  {2021})}\BibitemShut {NoStop}%
\bibitem [{\citenamefont {Madsen}\ \emph {et~al.}(2022)\citenamefont {Madsen},
  \citenamefont {Laudenbach}, \citenamefont {Askarani}, \citenamefont
  {Rortais}, \citenamefont {Vincent}, \citenamefont {Bulmer}, \citenamefont
  {Miatto}, \citenamefont {Neuhaus}, \citenamefont {Helt}, \citenamefont
  {Collins} \emph {et~al.}}]{madsen2022quantum}%
  \BibitemOpen
  \bibfield  {author} {\bibinfo {author} {\bibfnamefont {L.~S.}\ \bibnamefont
  {Madsen}}, \bibinfo {author} {\bibfnamefont {F.}~\bibnamefont {Laudenbach}},
  \bibinfo {author} {\bibfnamefont {M.~F.}\ \bibnamefont {Askarani}}, \bibinfo
  {author} {\bibfnamefont {F.}~\bibnamefont {Rortais}}, \bibinfo {author}
  {\bibfnamefont {T.}~\bibnamefont {Vincent}}, \bibinfo {author} {\bibfnamefont
  {J.~F.}\ \bibnamefont {Bulmer}}, \bibinfo {author} {\bibfnamefont {F.~M.}\
  \bibnamefont {Miatto}}, \bibinfo {author} {\bibfnamefont {L.}~\bibnamefont
  {Neuhaus}}, \bibinfo {author} {\bibfnamefont {L.~G.}\ \bibnamefont {Helt}},
  \bibinfo {author} {\bibfnamefont {M.~J.}\ \bibnamefont {Collins}}, \emph
  {et~al.},\ }\bibfield  {title} {\bibinfo {title} {Quantum computational
  advantage with a programmable photonic processor},\ }\href@noop {} {\bibfield
   {journal} {\bibinfo  {journal} {Nature}\ }\textbf {\bibinfo {volume}
  {606}},\ \bibinfo {pages} {75} (\bibinfo {year} {2022})}\BibitemShut
  {NoStop}%
\end{thebibliography}%

\end{document}